\documentclass[twocolumn,showpacs,showkeys,prd]{revtex4}
\usepackage{graphicx}

\def\meV{\hbox{ meV}}
\def\eV{\hbox{ eV}}
\def\GeV{\hbox{ GeV}}
\def\TeV{\hbox{ TeV}}

\begin{document}

\title{Majorana neutrino magnetic moments in the gauge mediated
supersymmetry breaking MSSM model}

\author{Marek G\'o\'zd\'z}
\email{mgozdz@kft.umcs.lublin.pl}
\author{Wies{\l}aw A. Kami\'nski} 
\email{kaminski@neuron.umcs.lublin.pl}

\affiliation{
Department of Informatics, Maria Curie-Sk{\l}odowska University, \\
pl. Marii Curie--Sk{\l}odowskiej 5, 20-031 Lublin, Poland}

\begin{abstract}
  Supersymmetric models with broken $R$-parity provide mechanisms that
  allow to generate Majorana neutrino masses and magnetic moments
  through virtual particle-sparticle loops. This constitutes an
  attractive alternative to the see-saw mechanism. In this paper we
  present a~detailed calculation of the transition magnetic moments of
  a~Majorana neutrino in gauge mediated supersymmetry breaking MSSM
  without $R$-parity. We base our analysis on the renormalization group
  evolution of the MSSM parameters, which are unified at the GUT scale.
\end{abstract}

\pacs{12.60.Jv, 11.30.Pb, 14.60.Pq} 
\keywords{neutrino magnetic moment, supersymmetry, R-parity, gauge
  mediated supersymmetry breaking}

\maketitle

\section{Introduction}

After establishing the fact that neutrinos do oscillate \cite{nu-osc},
the window to physics beyond the Standard Model (SM) has been opened. It
is difficult to guess to what extend the already known theory of
elementary particles and interactions needs altering. It is customary to
believe, however, that the SM should be treated as a~low-energy
approximation of a~more general theory, which will not only work for
high energies, thus describing the creation of the Universe, but should
also use a~unified description of all the interactions, presumably
including gravity. A~good candidate seems to be somehow connected with
the string theory, which in turn requires supersymmetry (as well as
additional spatial dimensions) for consistency.

A~close cooperation between the development of theory and experiments is
essential. Despite the fact that direct testing of these models in the
ultra-high energy regime is by now impossible, different models may
foresee certain features of some elementary particles, branching ratios
and others. These subtle clues, when found in future generation
experiments, may lead to favoring some and ruling out the other models,
providing an important insight into high-energy exotic physics. One
cannot therefore underestimate the importance of study of different
theories beyond the SM and their implications.

One of the most promising concepts that extends beyond the SM is
supersymmetry (SUSY). It is strongly connected with the string theory,
which in order to be able to describe not only interactions (bosonic
strings) but also matter (fermionic strings) requires the introduction
of SUSY. SUSY provides an elegant way of describing fermionic and
bosonic fields grouped in a~single supermultiplet, and it is a~basic
exercise to show that each supermultiplet must consist of equal number
of degrees of freedom of both kinds. Therefore, introduction of SUSY
unifies in some sense the description of matter and interactions. What
is more, the Minimal Supersymmetric Standard Model (MSSM; see
\cite{mssm,kazakov} and references therein for a~review) possesses the
attractive feature that the gauge couplings unify at the energy $m_{\rm
  GUT} \sim 10^{16}$ GeV, which is not true in the ordinary SM. It is
remarkable that in order to go beyond the SM in a~consistent way one is
forced to accept a~whole bunch of new ideas like supersymmetry, extra
dimensions, grand unification (GUT) and others. The problem, however, is
that nobody can really state the actual details of these models. For
example, supposing that supersymmetry exists, it needs to be broken, as
it is not observed in our energy regime. Of course the details of the
mechanism of this breaking are not known. The difficulty with extra
dimensions is that one needs to justify why they cannot be seen, why do
they not open, what is the mechanism of compactification and
stabilization. The pattern and mechanism of unification of matter and
interactions at $m_{\rm GUT}$ or $m_{\rm Planck}$ can also be only
a~guess.

As mentioned at the very beginning, the only link we directly
investigate, leading beyond the SM, are neutrinos. In spite of the fact
that it is a~neutral particle, in certain exotic models it may possess
non-zero transition magnetic moments (in the case of Majorana neutrinos
this is the only possible type of magnetic moment; the Dirac neutrinos
may possess the transition as well as the diagonal magnetic
moments). This happens in all supersymmetric models in which the
so-called $R$-parity is not conserved
\cite{barbier,aul83,valle,np,rbreaking}. In principle this feature
should leave a~clear signature, but the present sensitivities of the
experiments are at best five orders of magnitude to weak. The
observation of an electromagnetic interaction of the neutrino would be
a~breakthrough and may give us important information about details of
the exotic models.

The problem of generating Majorana neutrino mass and transition magnetic
moments in $R$-parity violating MSSM has been widely discussed in the
literature \cite{Haug,Bhatta,Abada,rpvneutrinos,mg,mg-art09,mg-art15}.
Many older approaches used certain simplifying assumptions about the
low-energy mass spectrum of the MSSM model. This has been corrected by
the use of GUT conditions and renormalization group equations (RGE)
\cite{mg,mg-art09,mg-art15}, which made the whole discussion dependent
on a few unification parameters only. Up to our best knowledge, all
calculations made so far used the supergravity mechanism of
supersymmetry breaking.

In this paper we present detailed calculations performed assuming the
gauge mediated supersymmetry breaking mechanism, for the whole allowed
parameter space. The paper is organized as follows. In the next section
we define the model, which is the minimal supersymmetric standard model
with gauge mediated supersymmetry breaking and not conserved
$R$-parity. In Section III we describe our procedure of obtaining the
low-energy spectrum of the model, together with different constraints we
impose on the results. Next, we discuss the Majorana neutrino transition
magnetic moments and present numerical results. A~short conclusion
follows at the end.

\section{RpV MSSM with gauge mediated supersymmetry breaking}

The Minimal Supersymmetric Standard Model \cite{mssm,kazakov} is
a~minimal extension of the usual SM which incorporates supersymmetry. It
implies that each particle gains a~superpartner with spin different by
1/2 unit. There is also an additional Higgs doublet introduced, in order
to assign masses to the up- and down-type particles. In result, the
number of particles in MSSM roughly doubles that of the SM.

In basic formulation of the MSSM one assumes ad-hoc the conservation of
the lepton and baryon numbers. This is achieved by the introduction of
an artificial symmetry called the $R$-parity. It is defined as
$R=(-1)^{3B+L+2S}$, where $B$ is the baryon number, $L$ the lepton
number, and $S$ the spin of the particle. The definition implies that
all ordinary SM particles have $R=+1$ and all their superpartners have
$R=-1$. In theories preserving $R$-parity the product of $R$ of all the
interacting particles in a~vertex of a~Feynman diagram must be equal to
$1$. It follows that a~SUSY particle must decay into another SUSY particle,
thus the lightest SUSY particle must remain stable and is considered
a~good candidate for the cold dark matter. In many models this particle
is the lightest neutralino, but sometimes the gluino takes its place. In
the case of gauge mediated supersymmetry breaking the lightest stable
SUSY particle is the gravitino.

The main motivation for the introduction of $R$-parity is the
conservation of $L$ and $B$ numbers. However, we already know that at
least the flavour lepton numbers $L_e$, $L_\mu$, and $L_\tau$ are not
conserved, as has been seen in the neutrino oscillation experiments.
There is also a~strong suspicion that at higher energies the full $L$
symmetry may not be exact. From formal theoretical point of view,
nothing motivates the rejection of interaction terms that do violate the
$R$-parity. This leads us to $R$-parity violating (RpV) models, which
exhibit richer and more interesting phenomenology. 

The full RpV MSSM model is described by the superpotential, which
includes the Lagrangian as its $F-$term. It consists of two parts:
$W=W^{MSSM} + W^{RpV}$. The $R$-parity conserving part of the
superpotential of MSSM is usually written as
\begin{eqnarray}
  W^{MSSM} &=& \epsilon_{ab} [
  (\mathbf{Y}_E)_{ij} L_i^a H_u^b \bar E_j
  + (\mathbf{Y}_D)_{ij} Q_{ix}^{a} H_d^b \bar D_{j}^{x} \nonumber \\
  &+& (\mathbf{Y}_U)_{ij} Q_{i x}^{a} H_u^b \bar U_{j}^{x} + \mu H_d^a H_u^b], 
\end{eqnarray}
while its RpV part reads
\begin{eqnarray}
  W^{RpV} &=& \epsilon_{ab}\left[
    \frac{1}{2} \lambda_{ijk} L_i^a L_j^b \bar E_k
    + \lambda'_{ijk} L_i^a Q_{jx}^{b} \bar D_{k}^{x} \right] \nonumber \\
  &+& \frac{1}{2}\epsilon_{xyz} \lambda''_{ijk}\bar U_i^x\bar
  D_j^y \bar D_k^z + \epsilon_{ab}\kappa^i L_i^a H_u^b.
\end{eqnarray}
The {\bf Y}'s are 3$\times$3 Yukawa matrices. $L$ and $Q$ are the
$SU(2)$ left-handed doublets while $\bar E$, $\bar U$ and $\bar D$
denote the right-handed lepton, up-quark and down-quark $SU(2)$
singlets, respectively. $H_d$ and $H_u$ mean two Higgs doublets. We have
introduced color indices $x,y,z = 1,2,3$, generation indices
$i,j,k=1,2,3=e,\mu,\tau$ and the $SU(2)$ spinor indices $a,b = 1,2$.

The mass terms (self-interaction terms) for the Higgs bosons, sfermions, and
gauginos take the standard form:
\begin{eqnarray}
{\cal L}^{mass} 
&=& \mathbf{m}^2_{H_d} h_d^\dagger h_d +
    \mathbf{m}^2_{H_u} h_u^\dagger h_u +
     q^\dagger \mathbf {m}^2_Q q + l^\dagger \mathbf {m}^2_L l 
\nonumber \\
&+&  u \mathbf {m}^2_U u^\dagger + d \mathbf {m}^2_D d^\dagger +
     e \mathbf {m}^2_E e^\dagger \\
  &+& \frac12 \left( 
  M_1 \tilde{B}^\dagger \tilde{B} + 
  M_2 \tilde{W_i}^\dagger \tilde{W^i} +
  M_3 \tilde{g_\alpha}^\dagger \tilde{g^\alpha} + h.c.\right ),
\nonumber
\end{eqnarray}
where the second part represents bino, wino, and gluinos
($\alpha=1,\dots,8$), and lower case letters denote the scalar part of
the respective superfield.

There are a~few schemes of supersymmetry breaking among which the two
most popular are the supergravity (SUGRA) and the gauge mediated (GMSB)
mechanisms. In SUGRA \cite{kazakov,sugra} the SUSY breaking occurs at
the Planck scale, so that no supersymmetry is observed in the whole
energy regime except the $m_{\rm Planck}$, where gravity enters the
game.

In the GMSB mechanism \cite{kazakov,gmsb} the scale of SUSY breaking is
much lower, and is defined by the characteristic scale of an
intermediate messenger sector. The assumption is, that SUSY is broken in
a~hidden (secluded) sector, whose detailed structure does not change the
phenomenology of the low-energy world. In our approach we assumed that
the secluded sector consists of a gauge singlet superfield $\hat S$,
whose lowest $S$ and $F$ components acquire vacuum expectation values
(vev).

\begin{figure}[ht]
  \includegraphics[width=0.3\textwidth]{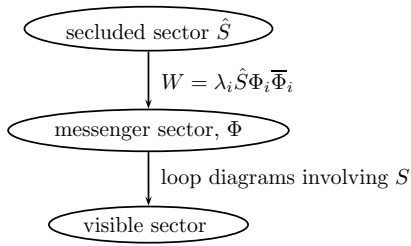}
  \caption{\label{fig:gmsb-schemat} The gauge mediated scheme of
    supersymmetry breaking (GMSB).}
\end{figure}

Supersymmetry breaking is communicated to the visible world via the
messenger sector (see Fig.~\ref{fig:gmsb-schemat}). The interaction
among superfields of the secluded and messenger sectors is described by
the superpotential
\begin{equation}
  W = \lambda_i \hat{S} \Phi_i {\overline \Phi_i}.
\end{equation}
where $\Phi_i$ and $\overline \Phi_i$ denote appropriate messenger
superfields. Because of nonzero vev of the lowest $S$ and $F$ components
of superfield $\hat S$, fermionic components of the messenger superfields
gain Dirac masses $M_i = \lambda_i S$ and determine in this way the
messenger scale $M$. Simultaneously mass matrices of their scalar
superpartners
\begin{equation}
 \left( \begin{array}{cc}
   |\lambda_i S|^2 & \lambda_i F \\
   \lambda_i^* F^* & |\lambda_i S|^2 
 \end{array} \right) 
\label{gm4}
\end{equation}
have eigenvalues $|\lambda_i S|^2 \pm |\lambda_i F|$.

It is easy to see that vev of $S$ generates masses for fermionic and
bosonic components of messenger superfields, while vev of $F$ destroys
degeneration of these masses, which results in supersymmetry
breaking. Defining $F_i \equiv \lambda_i F$ one can introduce a new
parameter $\Lambda_i \equiv F_i/S$ measuring the fermion--boson mass
splitting,
\begin{eqnarray}
  m_f &=& M_i, \nonumber \\
  m_b &=& M_i \sqrt{1 \pm \frac{\Lambda_i}{M_i}}.
\end{eqnarray}
Parameter $\Lambda$ and the messenger scale $M$ are in the following treated
as free parameters of the model.

Messenger superfields transmit SUSY breaking to the visible sector. It
is realized through loops containing insertions of $S$ and results in 
gaugino and scalar masses at $M$ scale:
\begin{equation}
  M_{\tilde \lambda_i}(M) = k_i \frac{\alpha_i(M)}{4 \pi}
  \Lambda_G,
  \label{gm5}
\end{equation}
\begin{equation}
  m^2_{\tilde f}(M) = 2 \sum_{i = 1}^{3} C_i^{\tilde f}k_i
  \left(\frac{\alpha_i(M)}{4 \pi}\right)^2\Lambda_S^2,
  \label{gm6}
\end{equation}
where  $i=1,2,3$ is the gauge group index, and
\begin{equation}
  \Lambda_G =\sum_{k = 1}^{N_g}n_k\frac{F_k}{M_k}
  g\left(\frac{F_k}{M_k^2}\right), 
\label{gm71}
\end{equation}
\begin{equation}
\Lambda_S^2 = \sum_{k = 1}^{N_g}n_k\frac{F_k}{M_k^2}
  f\left(\frac{F_k}{M_k^2}\right), 
\label{gm7}
\end{equation}
with $k$ being the flavor index. In Eqs.~(\ref{gm71}) and (\ref{gm7})
$n_k$ is the doubled Dynkin index of the messenger superfield
representation with flavor $k$. Coefficients $C_i^{\tilde f}$ are the
quadratic Casimir operators of sfermions. For $d$-dimensional
representation of $SU(d)$ their eigenvalues are $C = (d^2 - 1)/2d$. In the
case of $U(1)$ group, $C = Y^2 = (Q - T_3)^2$. It follows that coefficients
$k_i$ are equal to $5/3$, $1$, and $1$, for $SU(3)$, $SU(2)$, and $U(1)$,
respectively. The normalization here is conventional and assures that
all $k_i\alpha_i$ meet at the GUT scale. Finally, the functions $f$ and
$g$ have the following forms:
\begin{equation}
  g(x) = \frac{1}{x^2}[(1+x)\log(1+x)] + (x \to -x),
\end{equation}
\begin{eqnarray}
  f(x) &=& \frac{1 + x}{x^2}
  \bigg[\log(1 + x) - 2 Li_2\left(\frac{x}{1 + x}\right) \nonumber \\
    &+& {1 \over 2}Li_2\left(\frac{2x}{1 + x}\right)\bigg] + (x \to -x). 
\end{eqnarray}

In the minimal model of GMSB there is only one messenger field
flavor. Thus, dropping flavor indices, one can write Eqs.~(\ref{gm5})
and (\ref{gm6}), using the explicit forms Eqs.~(\ref{gm71}) and
(\ref{gm7}), as
\begin{equation}
  M_{\tilde \lambda_i}(M) = N k_i \frac{\alpha_i(M)}{4 \pi}
  \Lambda g\left(\frac{\Lambda}{M}\right), 
  \label{gm10}
\end{equation}
\begin{equation}
  m^2_{\tilde f}(M) = 2 N \sum_{i = 1}^{3} C_i^{\tilde f}k_i
  \left(\frac{\alpha_i(M)}{4\pi}\right)^2\Lambda^2
  f\left(\frac{\Lambda}{M}\right) \ {\bf 1},
\label{gm11}
\end{equation}
where $C_1^{\tilde f} = Y^2, C_2^{\tilde f} = 3/4$ for $SU(2)_L$
doublets and 0 for singlets, $C_3^{\tilde f}$ is equal to $4/3$ for
$SU(3)_C$ triplets and 0 for singlets. In Eq. (\ref{gm11}) ${\bf 1}$
denotes the unit matrix in generation space and guarantees the lack of
flavor mixing in soft breaking mass matrices at messenger scale. $N$,
the so-called generation index, is given by $N = \sum_{i = 1}^{N_g}n_i$,
where $N_g$ means the total number of generations. In this paper we
study the following two cases: (1) a single flavor of $5 + \overline 5$
representation of $SU(5)$, with $SU(2)_L$ doublets ($l$ and $\tilde l$)
and $SU(3)$ triplets ($q$ and $\tilde q$), and (2) a single flavor of
both representations $5 + \overline 5$ and $10 + \overline{10}$ of the
$SU(5)$ group. In case (1) $N$ is equal to 1, while in case (2) $N = 1 +
3 = 4$, because for $10 + \overline{10}$ representation of $SU(5)$ the
doubled Dynkin index is equal to 3.

\section{Obtaining and constraining the low-energy spectrum of the
  model}

The MSSM model has more than one hundred free parameters, which
drastically decreases its predictive power. The possible way out is to
use certain unification conditions at high energy scale $m_{\rm GUT}\sim
10^{16}$GeV and derive the low-energy values of all parameters by means
of the renormalization group equations. The set of free parameters can
in this way be reduced to few. This widely accepted approach connects
supersymmetry and grand unified theories, and is appropriate in the
SUGRA case. The main difference between SUGRA and GMSB is that in the
latter all the parameters are evolved between the weak scale $m_{\rm Z}$
and the messenger scale $M \ll m_{\rm GUT}$.  Besides, due to new
interactions with the messenger sector, the mass matrices are
constructed in a different way, which gives gravitino as the lightest
SUSY particle, and results in further corrections.

In our case the free parameters of the model are: $\Lambda$, the
splitting between fermion and boson masses, $M$, the characteristic
energy scale of the messenger sector, $\tan\beta \equiv v_u/v_d$, where
$v_u$ and $v_d$ are vevs of the $H_u$ and $H_d$ superfields, and
sgn$(\mu)$.

The whole procedure of obtaining the low-energy spectrum is explained in
great detail in Ref.~\cite{mg-gmsb} and here we will recall the basic
steps only. Everything starts with evolving all gauge and Yukawa
couplings up to the messenger scale $M$.  Despite the fact that the
heaviest third generation dominates, and it is customary to drop the
dependence on the remaining generations, we use all three of them in our
equations. For the RGE evolution the one-loop standard model equations
\cite{drtjones} are used below the mass threshold $M_{\rm SUSY}$, where
SUSY particles start to contribute, and the MSSM RGE \cite{martinvaughn}
above that scale. In our case the two-loop corrections, as well as
corrections coming from the RpV parts, can be safely neglected (for a
discussion of this problem see Ref.~\cite{mg-art1}). Initially, scale
$M_{\rm SUSY}$ is taken to be equal to 1~TeV, but this value is modified
during the running of the relevant masses. In the next step the gaugino
and sfermion soft mass matrices are constructed using Eqs.~(\ref{gm10})
and (\ref{gm11}), and the RGE evolution of all the quantities is
performed back to the $m_{\rm Z}$ scale. Meanwhile the electroweak
symmetry breaking (Higgs sector) is handled, which allows to obtain the
low-energy mass spectrum of the model.

Of course not all combinations of the values of the initial parameters
lead to a~physically acceptable mass spectrum. We test the obtained
results against four additional constraints, ie.: (1) finite values of
Yukawa couplings at the GUT scale; (2) proper treatment of the
electroweak symmetry breaking; (3) requirement of physically acceptable
mass eigenvalues at low energies; (4) FCNC phenomenology. The full
discussion of the allowed parameter range for our model, coming from
these constraints, is discussed in Ref.~\cite{mg-gmsb}.

\section{Majorana neutrino transition magnetic moments in GMSB MSSM}

\begin{figure*}
  \centering
  \includegraphics[width=0.40\textwidth]{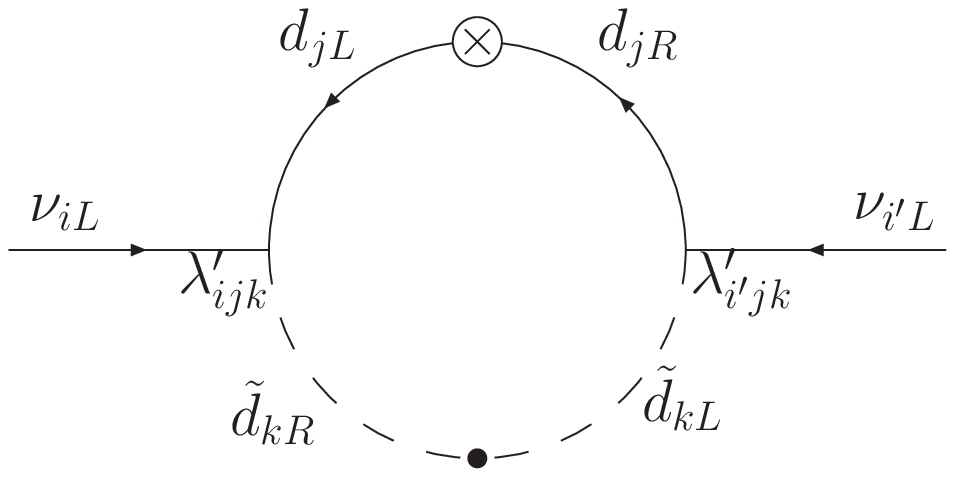} 
  \includegraphics[width=0.40\textwidth]{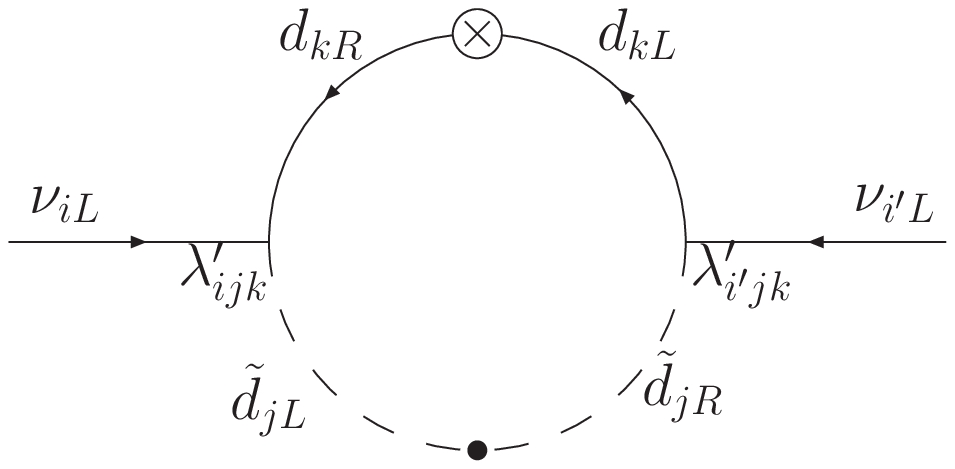} \\
  \includegraphics[width=0.40\textwidth]{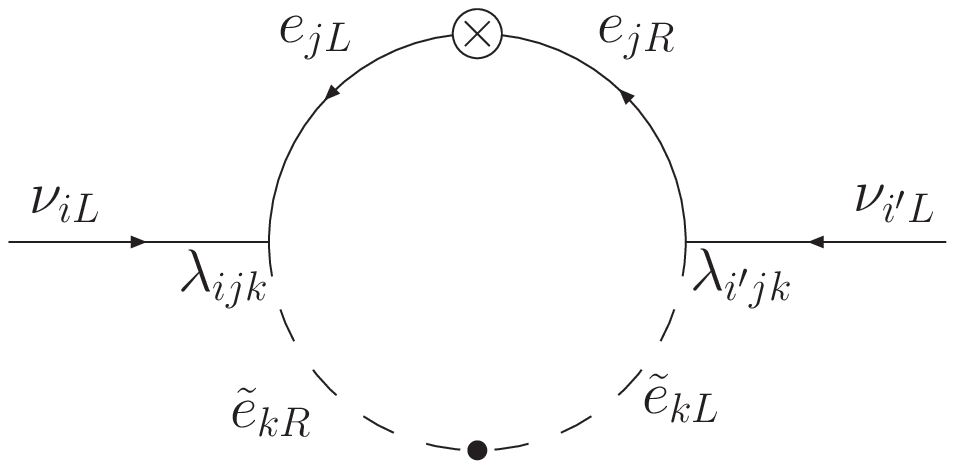} 
  \includegraphics[width=0.40\textwidth]{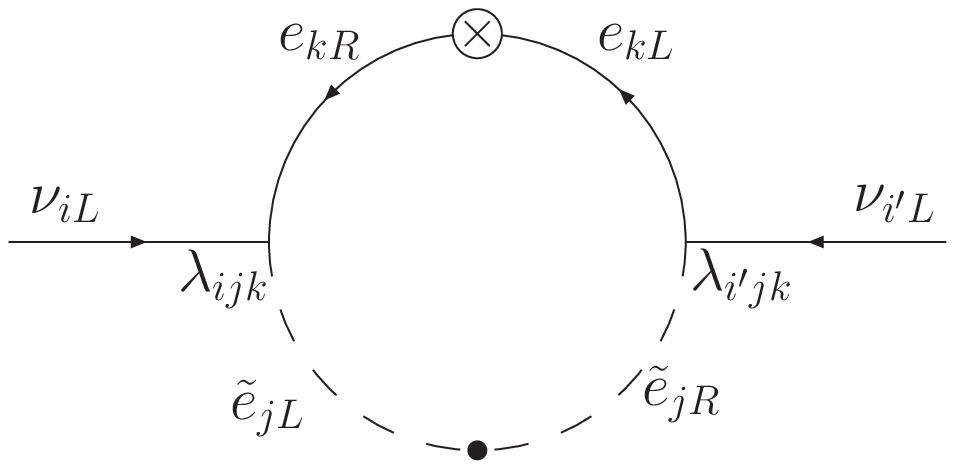} 
  \caption{\label{fig:loops} The basic 1--loop diagrams giving rise to
    the Majorana neutrino mass in the $R$-parity violating MSSM. The
    transition magnetic moment is obtained by attaching an external
    photon to the loops.}
\end{figure*}

The introduction of supersymmetry means doubling the number of particles
and introducing a~lot of new possible interactions among them. SUSY with
broken $R$-parity extends the possibility of exotic processes to occur.
It is well known, for example, that Majorana neutrinos may acquire
masses without the see-saw mechanism, due to one-loop processes in which
a~neutrino decays into a~particle-sparticle pair, which combines into
another neutrino of different flavour. The leading contributions to such
a~process are schematically depicted on Fig.~\ref{fig:loops}. In this
paper we consider two possibilities, with a quark and a~squark, and with
a~charged lepton and a~slepton inside the loop. Other contributions,
like the mixing of neutrinos with neutralinos, are much weaker
\cite{mg-art15} and are dropped here.

These processes effectively expand the neutrino--neutrino interaction
vertex into a~loop of virtual charged particles. This means that one may
attach an external photon to the loop; the amplitude of such interaction
would be proportional to the neutrino magnetic moment. The observation
of the electromagnetic interaction of a~neutrino will be a~strong
suggestion in favour of the RpV physics.

The problem of generating neutrino mass matrix from the RpV loops has
been extensively discussed in the literature
\cite{Haug,Bhatta,Abada,rpvneutrinos}, and various approaches and
approximations have been used by different authors. Our method
\cite{mg,mg-art09,mg-art15}, which involves the careful generation of
the low-energy spectrum of the model seems to be the most complete by
now. The calculation of the magnetic moments bases on the knowledge of
the neutrino mass matrix, and the latter may be obtained from the
experimental values of the mixing angles, under the assumption of
certain (normal or inverted) hierarchy of the neutrino masses.

The contribution to the magnetic moments coming from the squark--quark
loop reads \cite{mg-art09}:
\begin{eqnarray}
  \mu_{\nu_{ii'}}^q &=& (1-\delta_{ii'}) \frac{12 Q_d m_e}{16\pi^2}
  \sum_{jkl} \Bigg \{
  \lambda'_{ijk}\lambda'_{i'kl} \sum_a V_{ja} V_{la}
  \frac{w^q_{ak}}{m_{d^a}} \nonumber \\
  &-& \lambda'_{ijk}\lambda'_{i'lj} \sum_a V_{ka} V_{la}
  \frac{w^q_{aj}}{m_{d^a}} \Bigg \} \mu_B,
\label{Tqq}
\end{eqnarray}
where the loop integral $w$ takes the form
\begin{equation}
  w^q_{jk} = \frac{\sin 2\theta^k}{2} \left (
    \frac{x_2^{jk}\ln x_2^{jk}-x_2^{jk}+1}{(1-x_2^{jk})^2} -
    (x_2 \to x_1) \right ).
\end{equation}
Here $Q_d = 1/3$ is the $d$-quark charge in units of $e$, and $m_e$
denotes the electron mass. $V=V_{\rm CKM}$ is the
Cabibbo--Kobayashi--Maskawa quark mixing matrix, as we take into account
the fact that quarks may mix inside the loops. $\mu_B$ denotes the Bohr
magneton. We have defined dimensionless quantities $x_1^{jk} \equiv
m_{d^j}^2 / m_{\tilde d_1^k}^2$ and $x_2^{jk} \equiv m_{d^j}^2 /
m_{\tilde d_2^k}^2$ representing particle to sparticle mass ratios
squared.

In the case of the slepton--lepton loop two modifications are in
order. Firstly, the mixing of leptons is negligible,
secondly, leptons are colorless, so a~factor of three drops out from the
formula. We end up with:
\begin{eqnarray}
\label{Tll}
  \mu_{\nu_{ii'}}^\ell &=& (1-\delta_{ii'}) \frac{4 Q_e m_e}{16\pi^2} \\
  &\times& \sum_{jk}
  \lambda_{ijk}\lambda_{i'kj} \left(
    \frac{w^\ell_{jk}}{m_{e^j}} - \frac{w^\ell_{kj}}{m_{e^k}} \right ) 
  \mu_B, \nonumber
\end{eqnarray}
where the loop integral is equal to
\begin{equation}
  w^\ell_{jk} = \frac{\sin 2\phi^k}{2} \left (
    \frac{y_2^{jk}\ln y_2^{jk}-y_2^{jk}+1}{(1-y_2^{jk})^2} -
    (y_2 \to y_1) \right ).
\end{equation}
Again, we have defined dimensionless quantities $y_1^{jk} \equiv
m_{e^j}^2 / m_{\tilde e_1^k}^2$ and $y_2^{jk} \equiv m_{e^j}^2 /
m_{\tilde e_2^k}^2$.

As one can see, in order to calculate $\mu_\nu$ one needs to know the
RpV couplings $\lambda$ and $\lambda'$. These are in principle unknown
free parameters of the model but fortunately it is possible to get rid
of this obstacle by the use of the mass matrices. The latter may be
expressed as
\begin{eqnarray}
  {\cal M}_{ii'}^q &=&
  \frac{3}{16\pi^2} \sum_{jkl} \Bigg\{ 
  \left(
    \lambda'_{ijk}\lambda'_{i'kl} \sum_a V_{ja} V_{la} v^q_{ak} m_{d^a}
  \right ) \nonumber \\
  &+& \left (
    \lambda'_{ijk}\lambda'_{i'lj} \sum_a V_{ka} V_{la} v^q_{aj} m_{d^a}
  \right ) \Bigg \},
\label{Mqq}
\end{eqnarray}
\begin{equation}
  {\cal M}_{ii'}^\ell = 
  \frac{1}{16\pi^2} \sum_{jk} \lambda_{ijk}\lambda_{i'kj} 
  (v^\ell_{jk} m_{e^j} + v^\ell_{kj} m_{e^k}),
\label{Mll}
\end{equation}
with $v^{\ell,q}$ being another loop integrals \cite{mg-art09}. Now, we
assume that each mechanism (ie. each combination of indices labeling
$\lambda$ and $\lambda'$) may be analyzed separately. This is a~usual
approach, which is justified by the assumption that there is no
fine-tuning between different processes that contribute to $\cal M$. In
this convenient situation only one element from the sums in $\cal M$ is
present at a~time, thus reducing the expressions to a~much simpler
form. This allows one to substitute the unknown products
$\lambda\lambda$ and $\lambda'\lambda'$ in Eqs.~(\ref{Tqq}) and
(\ref{Tll}) by the respective mass matrix elements. The advantage of
such approach is obvious, as one may construct $\cal M$ numerically
using experimental data. 

Finally one gets for the magnetic moments (for more details see
Ref.~\cite{mg-art09})
\begin{eqnarray}
  \label{eq:muq}
  \mu_{\nu_{ii'}}^q &\simeq& 
  (1-\delta_{ii'}) {\cal M}^q_{ii'} f^q_{\rm SUSY}, \\
  \mu_{\nu_{ii'}}^\ell &\simeq& 
  (1-\delta_{ii'}) {\cal M}^\ell_{ii'} f^\ell_{\rm SUSY},
\end{eqnarray}
where the functions $f_{\rm SUSY}$ convert the neutrino masses into
magnetic moments and depend on the particles masses and V matrix
elements. Their explicit form and values for different SUSY input
parameters can be found in \cite{mg-art09}, but overall these are
numbers between roughly $0.5\times 10^{-15}$ and $2.7\times
10^{-18}$. The full transition magnetic moment would consist of both
contributions, ie.
\begin{equation}
  \mu_{\nu_{ii'}} = \mu_{\nu_{ii'}}^\ell + \mu_{\nu_{ii'}}^q.
\end{equation}

\begin{table}[t]
  \centering
  \caption{\label{tab:1}Lower and upper bounds on the Majorana neutrino
    transition magnetic moments in GMSB MSSM, assuming normal (NH) or
    inverted hierarchy (IH), and two different structures of the messenger
    sector with the generation index $N=1,4$. The whole allowed parameter space has been considered. Here,
    sgn$(\mu)=+1$. The unit is the Bohr magneton $\mu_B$. }
  \begin{tabular}{cccc}
    \hline\hline
    hier. & ~~~N~~~ & $\mu_{\nu_{e\mu}}$, $\mu_{\nu_{e\tau}}$ & $\mu_{\nu_{\mu\tau}}$ \\
    \hline
    NH & 1 &
    $(0.38, 28.3)\times 10^{-19}$~~~ & 
    $(0.28, 20.5)\times 10^{-18}$ \\
    NH &  4 &
    $(0.73, 66.1)\times 10^{-20}$~~~ & 
    $(0.53, 47.9)\times 10^{-19}$ \\
    IH &  1 &
    $(0.36, 26.2)\times 10^{-20}$~~~ & 
    $(0.33, 24.0)\times 10^{-18}$ \\
    IH &  4 &
    $(0.68, 61.3)\times 10^{-21}$~~~ & 
    $(0.62, 56.0)\times 10^{-19}$ \\
    \hline\hline
  \end{tabular}
\end{table}

\begin{table}[t]
  \centering
  \caption{\label{tab:2} Same as Tab.\ref{tab:1} but for sgn$(\mu)=-1$.}
  \begin{tabular}{cccc}
    \hline\hline
    hier. & ~~~N~~~ & $\mu_{\nu_{e\mu}}$, $\mu_{\nu_{e\tau}}$ & $\mu_{\nu_{\mu\tau}}$ \\
    \hline
    NH & 1 &
    $(0.39, 23.8) \times 10^{-19}$~~~ &
    $(0.28, 17.3) \times 10^{-18}$ \\
    NH & 4 &
    $(0.16, 6.67) \times 10^{-19}$~~~ &
    $(0.11, 4.84) \times 10^{-18}$ \\
    IH & 1 &
    $(0.36, 22.1) \times 10^{-20}$~~~ &
    $(0.33, 20.2) \times 10^{-18}$ \\
    IH & 4 &
    $(0.15, 6.19) \times 10^{-20}$~~~ &
    $(0.13, 5.65) \times 10^{-18}$ \\
    \hline\hline
  \end{tabular}
\end{table}

\begin{figure*}
  \centering
  \includegraphics[width=0.4\textwidth,angle=270]{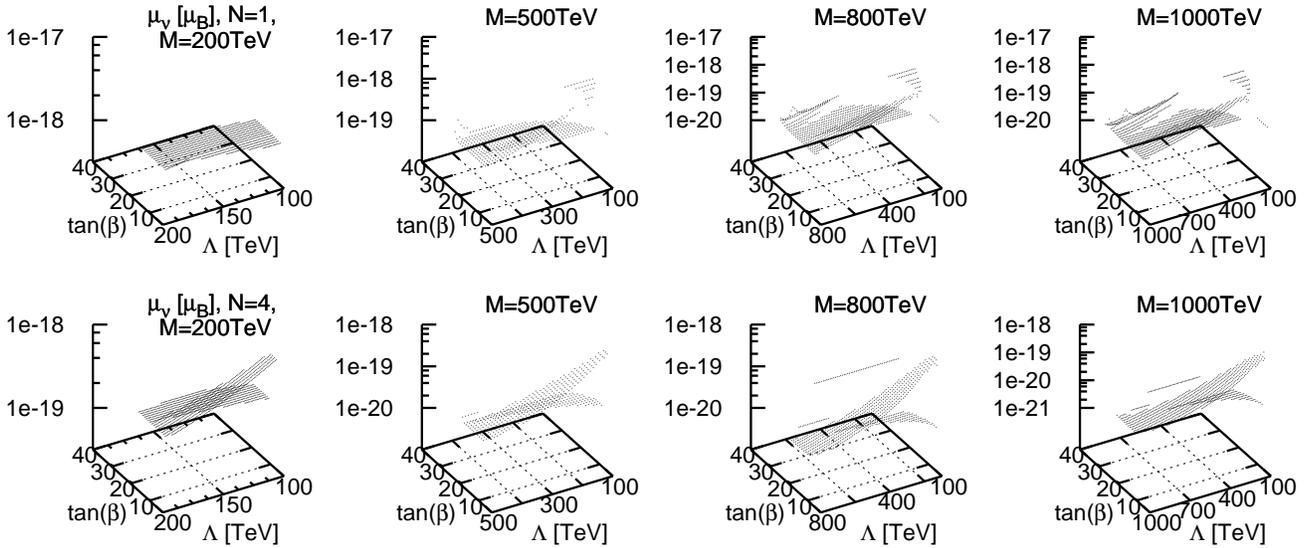}
  \caption{\label{fig:scan1} Neutrino magnetic moment $\mu_{\nu_{e\mu}}$
    for certain values of the GMSB parameters. Here, sgn$(\mu)=+1$ and
    normal hierarchy of neutrino masses is assumed.}
\end{figure*}

\begin{figure*}
  \centering
  \includegraphics[width=0.4\textwidth,angle=270]{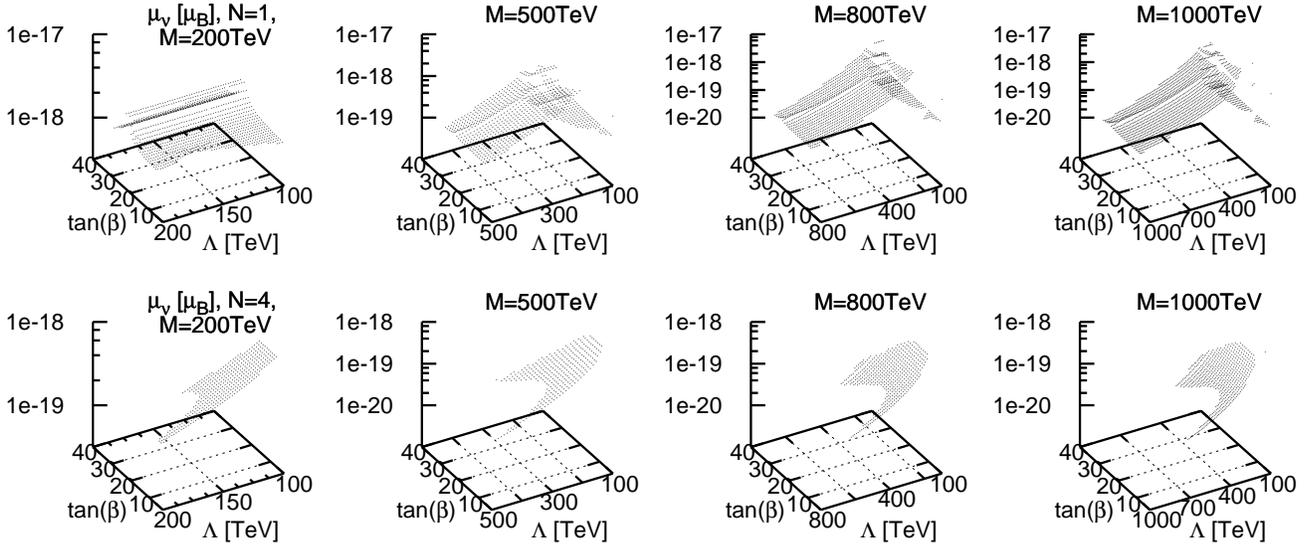}
  \caption{\label{fig:scan2} Same as Fig.~\ref{fig:scan1} but for
    sgn$(\mu)=-1$.}
\end{figure*}

We have calculated the transition magnetic moments $\mu_{\nu_{e\mu}}$,
$\mu_{\nu_{e\tau}}$, and $\mu_{\nu_{\mu\tau}}$ using the following values
of the input parameters:
\begin{eqnarray}
  && 3 \le \tan\beta \le 40, \\
  && 100 \TeV \le \Lambda < M,  \\
  && M = 200, 500, 800, 1000 \TeV, \\
  && \mathrm{sgn}(\mu) = \pm 1, \quad N=1,4.
\end{eqnarray}
The $\Lambda$ parameter was incremented by 1 for $M=200\TeV$, and by 10
for $M=500,800,1000\TeV$. $\tan\beta$ was incremented by 1. 

The construction of the neutrino mass matrix ${\cal M}$ is
straightforward. We use the standard trigonometric parameterization of
${\cal M}$ and the following values of the mass and mixing parameters
\cite{nu-osc,nu-mass}: $\Delta m_{12}^2 = 7.1 \times 10^{-5} \eV^2$,
$\Delta m_{23}^2 = 2.1 \times 10^{-3} \eV^2$, $\sin^2(\theta_{12}) =
0.2857$, $\sin^2(\theta_{23}) = 0.5$, $\sin^2(\theta_{13}) = 0$. As will
be seen later, the actual numbers chosen here are not essential. In the
matter of fact, one of the most recent analysis suggests the best-fit
value of the $\sin^2(\theta_{13})$ parameter to be slightly above zero
\cite{fogli}. However, in our case this change plays no role, as the
dominant part, which determines the overall order of magnitude of
$\mu_\nu$, is the $f_{\rm SUSY}$ function. Additionally, we assume that
the lightest neutrino mass is zero, and that the CP symmetry is
conserved, which eliminates all the phase dependencies. This results for
the normal hierarchy (NH) in
\begin{equation}
  {\cal M}^{\rm NH} = \pmatrix{
    2.41  &  2.69 &   2.69 \cr
    2.69  & 25.53 &  19.51 \cr
    2.69  & 19.51 &  25.53} \meV,
  \label{eq:MNH}
\end{equation}
and for the inverted hierarchy (IH) in
\begin{equation}
  {\cal M}^{\rm IH} = \pmatrix{
  45.27  &  0.25 &   0.25 \cr
   0.25  & 22.80 &  22.80 \cr
   0.25  & 22.80 &  22.80} \meV.
  \label{eq:MIH}
\end{equation}

Fig.~\ref{fig:scan1} presents values of the $\mu_{\nu_{e\mu}}$
transition magnetic moment for sgn$(\mu)=+1$ and normal hierarchy of the
neutrino masses. The non-rectangular shapes come from the constraints on
the low-energy spectrum, and the higher the value of $M$ is chosen, the
more steep the results are. For example, for $M\sim 1000\TeV$ the
difference between lowest and highest values of $\mu_\nu$ reaches three
orders of magnitude, while for small $M\sim 200 \TeV$ $\mu_\nu$ is
nearly constant. The dependence on $\Lambda$ is monotonic, but changes
its character for $\tan\beta$ equals roughly 25. For small $\tan\beta$
$\mu_\nu$ is an decreasing function of $\Lambda$, while for high
$\tan\beta$ it becomes an increasing function. The steepness of this
function, as was stated above, increases with $M$. The general behaviour
is that for small $\Lambda$ the dependence on $\tan\beta$ becomes
strong, while the values of $\mu_{\nu_{e\mu}}$ converge for higher
$\Lambda$ and become nearly insensitive on $\tan\beta$. The difference
between $N=1$ and $N=4$ is that for higher $N$ the overall order of
magnitude is decreased by one. Also the resulting mass spectrum is
different, so that the shapes in Fig.~\ref{fig:scan1} (lower row) are
more constrained, than those for $N=1$ (upper row).

A~similar plot for sgn$(\mu)=-1$ is presented on
Fig.~\ref{fig:scan2}. The change in the sign of the $\mu$ parameter
results in a~completely different behaviour of the magnetic moments as
functions of the input parameters. For $N=1$ there are two discontinued
regions, which separate roughly at $\Lambda\approx 200\TeV$. The remark
about monotonicity and its dependence on $\tan\beta$, which was visible
in the previous case, is valid also here, but to much weaker extend,
except the narrow region $\Lambda\approx 200\TeV$. Of course, for the
case $M=200\TeV$, for which $\Lambda<200\TeV$ (recall that always
$\Lambda < M$), this feature isn't present. So for sgn$(\mu)=-1$ and
$N=1$ the $\Lambda$ parameter dominates the change in behaviour of the
magnetic moments. When switching to $N=4$, the shapes become nearly
smooth surfaces. The dependence on $\tan\beta$ is quite weak, in
comparison with the previous cases, while the dependence on $\Lambda$ is
a~monotonic one with decreasing character. The $M$ parameter shows its
impact in the same way as for sgn$(\mu)=+1$, ie. it stretches the shapes
along the $\mu_\nu$ axis. The gain here is only one order of magnitude,
when comparing the cases $M=200\TeV$ and $M=1000\TeV$.

It is worth to notice, that the assumption of inverted hierarchy would
not change qualitatively the behaviour of $\mu_\nu$, and therefore we do
not include separate plots for this case. The only change would be an
overall shift of the results along the $\mu_\nu$ axis, according to
different values of the mass matrix elements for the NH and IH cases.

Also the remaining two transition magnetic moments, $\mu_{\nu_{e\tau}}$
and $\mu_{\nu_{\mu\tau}}$, exhibit very similar behaviour. The
$\mu_{\nu_{e\tau}}$ magnetic moment is to a~very good approximation
equal to $\mu_{\nu_{e\mu}}$, while the $\mu_{\nu_{\mu\tau}}$ will have
values shifted up by roughly one order of magnitude (see below).

A~summary of the upper and lower limits of the magnetic moments for all
considered combinations of the input parameters are presented in
Tabs.~\ref{tab:1} and \ref{tab:2}. In most cases, they span over
two-three orders of magnitude. There is also a~general trend that
$\mu_{\nu_{\mu\tau}}$ has a~factor of 10 higher values than
$\mu_{\nu_{e\mu}}\approx\mu_{\nu_{e\tau}}$, which comes from the fact
that respective mass matrix elements scale in the same way
[cf. Eqs.~(\ref{eq:MNH}) and (\ref{eq:MIH})].

\section{Conclusions}

In the present paper we have used the gauge mediated supersymmetry
breaking version of the minimal supersymmetric standard model without
$R$-parity to calculate Majorana neutrino transition magnetic moments.
In order to reduce the number of free parameters, we have assumed a~GUT
unification at high energy scale $m_{\rm GUT}\sim 10^{16}\GeV$, and then
used the RGE equations to render the values of mass parameters and
coupling constants to the low-energy regime.

The magnetic moments are in our approach dependent on the choice of the
following parameters: $\Lambda$, $M$, $N$, $\tan\beta$, sgn$(\mu)$, and
the phenomenological neutrino mass matrix ${\cal M}$. The latter can be
calculated using the mixing parameters extracted from experiments,
assuming normal or inverted pattern of neutrino mass hierarchy.

We have discovered that the weakest dependence of $\mu_\nu$ comes from
the ${\cal M}$ matrix, which enters the formulas (\ref{eq:muq}) as
a~simple multiplicative factor. The dependence on $\Lambda$, $M$, $N$,
and $\tan\beta$ is rather complicated and difficult to describe. It is
presented on Figs.~\ref{fig:scan1} and \ref{fig:scan2}. A~substantial
qualitative change in the behaviour of $\mu_\nu$ can be observed when
the sign of the $\mu$ parameter is changed. In general, while for
sgn$(\mu)=+1$ the small and large values of $\tan\beta$ changed
qualitatively the behaviour of $\mu_\nu$, such a~collapse for
sgn$(\mu)=-1$ is driven by the $\Lambda$ parameter.

This all shows, that even if the neutrino magnetic moment would be
observed in an experiment, in most cases it will not allow to state
definite conclusions about the values of the paramaters in the context
of the discussed model. With some luck, it may, however, serve as a~clue
about the neutrino mass hierarchy, if it happens to place in a~region
covered by only one range listed in Tabs.~\ref{tab:1} and
\ref{tab:2}.


\section*{Acknowledgments}

The first author (MG) acknowledges the financial support from the Polish
State Committee for Scientific Research.


\end{document}